\documentclass{article}
\def\Ref#1{(\ref{#1})}
\usepackage{amsmath}
\usepackage{amssymb}
\usepackage{cite}
\def\d{{\rm d}}
\begin{document}
\begin{titlepage}
\noindent{\large\textbf{Solvable multi-species extensions of the
drop-push model}}

\vskip 2 cm

\begin{center}{Farinaz~Roshani$^{a,b}${\footnote
{farinaz@iasbs.ac.ir}} \& Mohammad~Khorrami$^a${\footnote
{mamwad@iasbs.ac.ir}}} \vskip 5 mm

\textit{ $^a$ Institute for Advanced Studies in Basic Sciences,
             P.~O.~Box 159,\\ Zanjan 45195, Iran. }

\textit{ $^b$ Institute for Studies in Theoretical Physics and
Mathematics, P.~O.~Box 5531,
             Tehran 19395, Iran. }
\end{center}

\begin{abstract}
\noindent A family of multispecies drop-push system on a
one-dimensional lattice is investigated. It is shown that this
family is solvable in the sense of the Bethe ansatz, provided a
nonspectral matrix equation is satisfied. The large-time behavior
of the conditional probabilities, and the dynamics of the
particle-type change are also investigated.
\end{abstract}
\begin{center} {\textbf{PACS numbers:}} 05.40.-a, 02.50.Ga

{\textbf{Keywords:}} reaction-diffusion, two-point function,
solvable, Bethe ansatz
\end{center}
\end{titlepage}
\section{Introduction}
Various aspects of one-dimensional asymmetric exclusion processes
have been of physical interest. These contain, for example, the
kinetics of biopolimerization \cite{bio}, dynamical models of
interface growth \cite{dy}, and the traffic models \cite{Na}. This
model is also related to the noisy Burgers equation \cite{Bur},
and hence to the study of shocks \cite{Der,Fer}. The dynamical
properties of this model have also been extensively studied, for
example in \cite{Fer,Lig,Gav}.

In the study of stochastic processes, the term solvable has been
used in several meanings. In
\cite{GSc,AKK,RK,AKK2,AA,RK3,RK2,RK4}, solvability means
solvability in the sense of the Bethe ansatz, or factorization of
the $N$-particle scattering matrix to the two-particle scattering
matrices. This is related to the fact that for systems solvable in
this sense, there are a large number of conserved quantities. In
\cite{BDb,BDb1,BDb2,BDb3,Mb,HH,AKA,KAA,MB,AAK}, solvability means
closedness of the evolution equation of the empty intervals (or
their generalization). And in \cite{GS,AAMS,SAK}, solvability
means that the evolution equation for the $n$-point functions,
contain only $n$- or less- point functions.

In \cite{GSc}, the Bethe ansatz was used to solve the asymmetric
simple exclusion process on a one-dimensional lattice. In
\cite{AKK}, a similar technique was used to solve the drop-push
model \cite{SD}, and a one-parameter family of reactions
containing the simple exclusion processes and the drop-push model
as special cases. In \cite{AKK2}, the same technique was used to
solve a two-parameter family of processes, involving bidirectional
diffusion, exclusion, and pushing. The behavior of this last model
on the continuum, was investigated in \cite{RK}. In all of the
above cases, the essence of the method has been to replace the
reactions by suitable boundary conditions. The boundary conditions
involved were extended in \cite{RK2} to describe systems in them
the process of annihilation exists as well, and in \cite{RK4}, to
general boundary conditions in the continuum.

All of the above studies have been about single-species systems.
In \cite{AA,RK3}, systems with exclusion processes were
investigated, which contained more than one species. It was shown
in \cite{AA}, that in order that such systems be solvable in the
sense of the Bethe ansatz, certain relations should be satisfied
between the rates. This relations can be written as some kind of a
spectral Yang-Baxter equation. In \cite{RK3}, it was shown that
this spectral equation is equivalent to a nonspectral matrix
equation involving the rates.

Here we want to extend this approach to the case of drop-push
models. The systems under consideration, consist of $N$ particles,
which can be of several species. This particles live on an
infinite one-dimensional lattice, so that each site of the lattice
is either empty or contains one particle. Each particle hops to
the site at its right-hand side with unit rate, if that site is
empty. If that site is occupied, then the particle may push the
other particle, and at the same time a reaction may occur between
these neighboring particles changing their species.

The scheme of the paper is the following. In section 2, a
multi-species extension of the drop-push model is introduced and
the use of a suitable boundary condition instead of the reaction
is investigated. In section 3, the solvability condition (in the
sense of the Bethe-ansatz) for this reaction is investigated, and
it is shown that this condition, which is a spectral equation for
the matrix of the reaction rates, can be rewritten as a
nonspectral equation for the same matrix. In section 4, the
conditional probability is obtained, and its behavior in the
two-particle sector, specially its large-time behavior, is
investigated. In section 5, the dynamics of the particle-type
change in the two-particle sector is investigated and the
large-time limit of the probability of particle-types is obtained.
Finally, section 6 is devoted to the concluding remarks.
\section{Multi-species extension of the drop-push model}
In the ordinary drop-push model, the system consists of a single
type of particles, living on a one-dimensional lattice. Each site
of the lattice is either empty or occupied by one particle. Any
particle can hop to the site at its right neighbor, with the rate
1, if that site is empty. If the right neighbor site is occupied,
the particle can still hop to that site and push the second
particle, with the same rate 1. One can write the reactions like
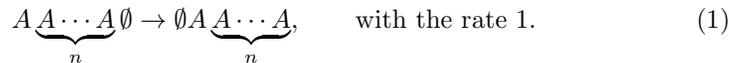
\begin{equation}\label{1}
A\underbrace{A\cdots A}_n\emptyset \to \emptyset
A\underbrace{A\cdots A}_n,\qquad \hbox{with the rate 1}.
\end{equation}
In this reaction, the total number of particles is conserved. For
a system containing $N$ particles, the question of interest is to
determine the probability of finding the $N$ particles in sites
$x_1$ to $x_N$, where
\begin{equation}\label{2}
x_i<x_j,\qquad\hbox{for }i<j,
\end{equation}
the so called physical region. It is easily seen that the
evolution equation for this probability, $P(x_1,\dots,x_N;t)$, is
\begin{align}\label{3}
\dot P(x_1,\dots,x_N;t)=&P(x_1-1,\dots,x_N;t)+\cdots+
P(x_1,\dots,x_N-1;t)\nonumber\\
&-N\,P(x_1,\dots,x_N;t),
\end{align}
if among the sites $x_i$, no two are adjacent; that is, if
$x_i<x_{i+1}-1$. For a block of $(n+1)$ adjacent sites, the
evolution equation becomes
\begin{align}\label{4}
\dot P(x_0=x,\dots,x_n=x+n;t)=&P(x_0=x-1,\dots,x_n=x+n;t)+
\cdots\nonumber\\
&+P(x_0=x-1,\dots,x_k=x+k-1,\nonumber\\
&\qquad\;\; x_{k+1}=x+k+1,\dots,x_n=x+n;t)\nonumber\\
&+\cdots -(n+1)P(x_0\dots,x_n;t).
\end{align}
This looks like different from \Ref{3}. However, defining a
boundary condition
\begin{equation}\label{5}
P(\dots,x,x,\dots):=P(\dots,x-1,x,\dots),
\end{equation}
makes the forms of \Ref{4} and \Ref{3} similar. One notes that
$P(\dots,x,x,\dots)$ has in fact no physical meaning, since the
argument of $P$ in that expression is not in the physical region.
But its introduction helps to solve the evolution equation, as it
was done in \cite{AAK} (and in \cite{GSc} for the exclusion
process).

Now suppose that the system consists of $k$ species of particles;
that is, an occupied site may have $k$ different states. Assume
moreover, that if the right neighbor of a particle is free, the
reaction is the same as ordinary drop-push model, without changing
the type of the particle, but there is a difference when two
particles are adjacent to each other: the left particle does push
the right one with unit rate, but in the mean time there is a
probability that the types of the particles change. So we have
reactions like
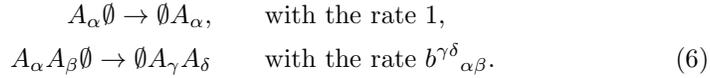
\begin{align}\label{6}
A_\alpha\emptyset\to\emptyset A_\alpha,\qquad& \hbox{with the rate
}1,\nonumber\\
A_\alpha A_\beta\emptyset\to\emptyset A_\gamma A_\delta\qquad&
\hbox{with the rate }b^{\gamma\delta}{}_{\alpha\beta}.
\end{align}
Consider a consisting of two particles, and denote the probability
that the first particle be at the site $x$ and of the type
$A_\alpha$ and the second particle be at the site $y$ and of the
type $A_\beta$, by $P^{\alpha\beta}(x,y)$. The evolution equations
become
\begin{align}\label{7}
\dot P^{\alpha\beta}(x,y;t)=&P^{\alpha\beta}(x-1,y;t)+
P^{\alpha\beta}(x,y-1;t)-2 P^{\alpha\beta}(x,y;t),\quad\!\!\!
x<y-1,
\nonumber\\
\dot P^{\alpha\beta}(x,x+1;t)=&P^{\alpha\beta}(x-1,x+1;t)+
b^{\alpha\beta}{}_{\gamma\delta}\,P^{\gamma\delta}(x-1,x;t) \nonumber\\
&-B^{\alpha\beta}\,P^{\alpha\beta}(x,x+1;t)-P^{\alpha\beta}(x,x+1;t),
\end{align}
where
\begin{equation}\label{8}
B^{\alpha\beta}:=\sum_{\gamma\delta}b^{\gamma\delta}{}_{\alpha\beta}.
\end{equation}
In fact, $B^{\alpha\beta}$ is the overall pushing rate, in which
the type change is unimportant. If this overall pushing rate is 1,
the second equation in \Ref{7} is simplified and it is seen that
it can be rewritten in the form of the first equation, provided
one introduces the boundary condition
\begin{equation}\label{9}
P^{\alpha\beta}(x,x;t)=b^{\alpha\beta}{}_{\gamma\delta}\,
P^{\gamma\delta}(x-1,x;t).
\end{equation}
One notes that all of the elements of the matrix $b$ (including
the diagonal elements) are nonnegative, as they are rates, and $b$
satisfies
\begin{equation}\label{10}
(\mathbf{s}\otimes\mathbf{s})b=\mathbf{s}\otimes\mathbf{s},
\end{equation}
where
\begin{equation}\label{11}
s_\alpha:=1.
\end{equation}
(This simply means that the sum of the elements each of the
columns of $b$ is equal to one.) A similar matrix $b$ was
introduced in \cite{RK3}, however the diagonal elements of that
$b$ were not necessarily nonnegative.

Now consider a system consisting of $N$ particles of various
species, with the evolution equation
\begin{align}\label{12}
\dot{\mathbf{P}}(x_1,\dots,x_N;t)=&\mathbf{P}(x_1-1,\dots,x_N;t)
+\cdots+ \mathbf{P}(x_1,\dots,x_N-1;t)\nonumber\\
&-N\,\mathbf{P}(x_1,\dots,x_N;t),
\end{align}
in the whole physical region, and the boundary condition
\begin{equation}\label{13}
\mathbf{P}(\dots,x_k=x,x_{k+1}=x,\dots):= b_{k,k+1}
\,\mathbf{P}(\dots,x_k=x-1,x_{k+1}=x,\dots),
\end{equation}
where
\begin{equation}\label{14}
b_{k,k+1}:=1\otimes\cdots\otimes 1\otimes\underbrace{b}_{k,k+1}
\otimes 1\otimes \cdots\otimes 1,
\end{equation}
and $\mathbf{P}$ is an $N$-tensor the components of which are
probabilities. It is seen that in this system, apart from the
simple diffusion, there is a reaction between a block of $n+1$
adjacent particles:
\begin{equation}\label{15}
A_{\alpha_0}\cdots A_{\alpha_n}\emptyset\to \emptyset
A_{\gamma_0}\cdots A_{\gamma_n},\qquad\hbox{with the rate }
(b_{n-1,n}\cdots b_{0,1})^
{\gamma_0\cdots\gamma_n}{}_{\alpha_0\cdots\alpha_n}.
\end{equation}
This comes from the fact that
\begin{align}\label{16}
&\mathbf{P}(x_0=x,\dots,x_{n-1}=x+n-1,x_n=x+n-1)\nonumber\\
=&(b_{n-1,n}\cdots
b_{0,1})\mathbf{P}(x_0=x-1,\dots,x_{n-1}=x+n-2,x_n=x+n-1).
\end{align}
Note the order of the matrices $b$. This order suggests that if a
collection of $n+1$ particles are adjacent, there is a probability
that the first particle pushes the second and changes the type of
the second (and itself) and then it is the second (modified)
particle that interacts with the third.
\section{Solvability and the Bethe-ansatz solution}
Consider the evolution equation \Ref{12} with the boundary
condition \Ref{13}. To solve this equation, one as usual seeks the
eigenvectors of the operator acting at the right-hand side of
\Ref{12}, that is, one tries to solve
\begin{align}\label{17}
E\,\mathbf{\Psi}(x_1,\dots,x_N)=&\mathbf{\Psi}(x_1-1,\dots,x_N)
+\cdots+ \mathbf{\Psi}(x_1,\dots,x_N-1)\nonumber\\
&-N\,\mathbf{\Psi}(x_1,\dots,x_N),
\end{align}
with
\begin{equation}\label{18}
\mathbf{\Psi}(\dots,x_k=x,x_{k+1}=x,\dots):= b_{k,k+1}
\,\mathbf{\Psi}(\dots,x_k=x-1,x_{k+1}=x,\dots).
\end{equation}
The Bethe-ansatz solution to this equation is
\begin{equation}\label{19}
\mathbf{\Psi}(\vec{x})=\sum_{\sigma}A_\sigma\,
e^{i\sigma(\vec{p})\cdot\vec{x}}\,\mathbf{\Xi},
\end{equation}
where $\mathbf{\Xi}$ is an arbitrary vector and the summation runs
over the elements of the permutation group of $N$ objects. Putting
this is \Ref{17}, one arrives at
\begin{equation}\label{20}
E=\sum_{k=1}^N(e^{-i p_k}-1),
\end{equation}
while \Ref{18} gives
\begin{equation}\label{21}
[1-e^{-i\sigma(p_k)}\, b_{k,k+1}]A_\sigma+
[1-e^{-i\sigma(p_{k+1})}\, b_{k,k+1}]A_{\sigma\sigma_k}=0,
\end{equation}
where
\begin{equation}\label{22}
\sigma_k(p_j)=\begin{cases}
                           p_{k+1},& j=k\\
                           p_k,& j=k+1\\
                           p_j,& j\ne k,k+1
               \end{cases}.
\end{equation}
From \Ref{21}, one arrives at
\begin{equation}\label{23}
A_{\sigma\sigma_k}=S_{k,k+1}[\sigma(p_k),\sigma(p_{k+1})]A_\sigma,
\end{equation}
where
\begin{equation}\label{24}
S_{k,l}(p_i,p_j):=-(1-z_j\, b_{k,l})^{-1}(1-z_i\, b_{k,l}),
\end{equation}
and
\begin{equation}\label{25}
z_j:=e^{-ip_j}.
\end{equation}
So one can construct $A_\sigma$'s from $A_1$, by writing $\sigma$
as a product of $\sigma_k$'s; that is, one can write the
$N$-particle scattering matrix $A$, as a product of the
two-particle scattering matrix $S$. However, as the generators of
the permutation group satisfy
\begin{equation}\label{26}
\sigma_k\sigma_{k+1}\sigma_k=\sigma_{k+1}\sigma_k\sigma_{k+1},
\end{equation}
one also needs
\begin{equation}\label{27}
A_{\sigma_k\sigma_{k+1}\sigma_k}=
A_{\sigma_{k+1}\sigma_k\sigma_{k+1}}.
\end{equation}
This, in terms of the $S$ becomes
\begin{equation}\label{28}
S_{1,2}(p_2,p_3)\, S_{2,3}(p_1,p_3)\, S_{1,2}(p_1,p_2)=
S_{2,3}(p_1,p_2)\, S_{1,2}(p_1,p_3)\, S_{2,3}(p_2,p_3).
\end{equation}
In terms of the $R$-matrix defined through
\begin{equation}\label{29}
S_{k,k+1}=:\Pi_{k,k+1}\, R_{k,k+1},
\end{equation}
\Ref{28} becomes
\begin{equation}\label{30}
R_{2,3}(p_2,p_3)\, R_{1,3}(p_1,p_3)\, R_{1,2}(p_1,p_2)=
R_{1,2}(p_1,p_2)\, R_{1,3}(p_1,p_3)\, R_{2,3}(p_2,p_3),
\end{equation}
which is the spectral Yang-Baxter equation.

The Bethe-ansatz solution exists, iff the scattering matrix
satisfies \Ref{28}. This is a general statement not coming from a
specific reaction. In the problem studied here, $S$ is of the form
\Ref{24}. Comparing this with the $S$-matrix obtained for the
multi-species simple exclusion process \cite{RK3}, it is seen that
$S(p_i,p_j)$ in these two different problems are transformed to
each other by a simple change $z_i\leftrightarrow z_j$. The
definition of $z$ in terms of $p$ in the present paper is,
however, different from that of \cite{RK3}. It is also seen that
with the changes $b_{1,2}\leftrightarrow b_{2,3}$ and
$z_1\leftrightarrow z_3$, \Ref{28} is transformed to the spectral
equation for $S$ obtained in \cite{RK3}. So it is not strange that
the \Ref{28} should be identical to another nonspectral equation.
For completeness, the argument leading to that nonspectral
equation is outlined. One notices that \Ref{28} is quadratic in
terms of $z_1$, and becomes identity when $z_1=z_2$ or $z_1=z_3$.
So \Ref{28} can be written as
\begin{equation}\label{31}
(z_1-z_2)(z_1-z_3) Q(z_2,z_3)=0,
\end{equation}
which means \Ref{28} is equivalent to $Q=0$. To obtain $Q$, one
can simply put $z_1=0$ in \Ref{28}. Doing this, and inverting both
sides, another equation is obtained which is quadratic in terms of
$z_3$. Again it is seen that for $z_3=0$ and $z_3=z_2$, this
equation becomes identity. So one can write this equation as
\begin{equation}\label{32}
z_3(z_3-z_2)\tilde Q(z_2)=0,
\end{equation}
which is equivalent to $\tilde Q=0$. To find $\tilde Q$, one
simply writes the coefficient of $z_3^2$ in the inverted equation.
One arrives at an equation containing only $z_2$. This in turn,
can be converted to an expression quadratic in terms of $z_2$. The
coefficients of $z_2^0$ and $z_2^2$ of this equation are
identities, while the coefficient of $z_2$ gives
\begin{equation}\label{33}
b_{2,3}\, b_{1,2}(b_{2,3}+b_{1,2})=(b_{2,3}+b_{1,2})b_{2,3}\,
b_{1,2},
\end{equation}
or
\begin{equation}\label{34}
b_{2,3}[b_{2,3},b_{1,2}]=[b_{2,3},b_{1,2}]b_{1,2},
\end{equation}
which are the same as eqs. (47) and (48) in \cite{RK3}, with
$b_{1,2}\leftrightarrow b_{2,3}$, as expected.
\section{The conditional probability}
Assuming that the solvability condition \Ref{28}, or equivalently
\Ref{34}, is satisfied, One can determine the conditional
probability (or the propagator):
\begin{equation}\label{35}
U(\vec{x};t|\vec{y};0)=\int\frac{\d^N p}{(2\pi)^N}
e^{-i\vec{p}\cdot\vec{y}}\sum_\sigma A_\sigma\,
e^{i\sigma(\vec{p})\cdot\vec{x}}\,e^{t\, E(\vec{p})},
\end{equation}
where the integration region for each $p_i$ is $[0,2\pi]$, and
$A_1=1$. The singularity in $A_\sigma$ is removed by setting
$p_j\to p_j-i\epsilon$, where the limit $\epsilon\to 0^+$ is
meant. Note that as the elements of $b$ are nonnegative and $b$
satisfies \Ref{10}, the absolute value of the eigenvalues of $b$
is not greater than 1.

For the two-particle sector, there is only one matrix ($b$) in the
expression of $U$. So, it can be treated as a $c$ number. Using a
calculation similar to what has been done in \cite{RK2} and
\cite{RK3}, one arrives at
\begin{align}\label{36}
U(\vec{x};t|\vec{y};0)=e^{-2t}\bigg\{&
\frac{t^{x_1-y_1}}{(x_1-y_1)!}\frac{t^{x_2-y_2}}{(x_2-y_2)!}\nonumber\\
+&\sum_{l=0}^\infty \frac{t^{x_2-y_1}}{(x_2-y_1)!}
\frac{t^{x_1-y_2-l}}{(x_1-y_2-l)!}
b^l\left[-1+\frac{(x_2-y_1)b}{t}\right]\bigg\}.
\end{align}

To investigate the large-time behavior of the propagator, it is
useful to decompose the vector space on which $b$ acts, into two
subspaces invariant under the action of $b$: the first subspace
corresponding to eigenvalues with modulus one, the second
corresponding to eigenvalues with modulus less than one. This is
done introducing two projections $Q$ and $R$, satisfying
\begin{align}\label{37}
Q+R&=1,\nonumber\\
Q\,R=R\,Q&=0,\nonumber\\
[b,Q]=[b,R]&=0.
\end{align}
$Q$ projects on the first subspace, and $R$ projects on the
second. One can now multiply $U$ by $1=Q+R$. In the term
multiplied by $R$, one can treat $b$ as a number with modulus less
than 1. But if the modulus of $b$ is less than 1, then the
integrand in \Ref{35} is nonsingular and it is readily seen that
for large $t$, the leading term in the integral is independent of
$b$, and in fact equal to the value obtained with $b=0$. So
\begin{align}\label{38}
U(\vec{x};t|\vec{y};0)=e^{-2t}\bigg\{&
\frac{t^{x_1-y_1}}{(x_1-y_1)!}\frac{t^{x_2-y_2}}{(x_2-y_2)!}\nonumber\\
+&\sum_{l=0}^\infty \frac{t^{x_2-y_1}}{(x_2-y_1)!}
\frac{t^{x_1-y_2-l}}{(x_1-y_2-l)!}
b^l\left[-1+\frac{(x_2-y_1)b}{t}\right]\bigg\}Q\nonumber\\
+e^{-2t}\bigg\{&
\frac{t^{x_1-y_1}}{(x_1-y_1)!}\frac{t^{x_2-y_2}}{(x_2-y_2)!}\nonumber\\
+&\sum_{l=0}^\infty \frac{t^{x_2-y_1}}{(x_2-y_1)!}
\frac{t^{x_1-y_2-l}}{(x_1-y_2-l)!}
b^l\left[-1+\frac{(x_2-y_1)b}{t}\right]\bigg\}R,
\end{align}
and for large times,
\begin{align}\label{39}
\hbox{the second term}=\frac{1}{2\pi t}\{&
e^{-[(x_1-y_1-t)^2+(x_2-y_2-t)^2]/(2t)}\nonumber\\
-&e^{-[(x_1-y_2-t)^2+(x_2-y_1-t)^2]/(2t)}\}R,\quad{t\to\infty}.
\end{align}
So at large times, the second term tends to zero faster than
$t^{-1}$, and the leading term in the conditional probability,
which is of the order $t^{-1}$, does not involve the second term.

If the only eigenvalue of $b$ with modulus 1 is 1, then $U$ has a
simple behavior for $t\to\infty$. In this case, $b\, Q=Q$, and one
can simplify $U$ to find
\begin{align}\label{40}
U(\vec{x};t|\vec{y};0)=e^{-2t}\bigg[&
\frac{t^{x_1-y_1}}{(x_1-y_1)!}\frac{t^{x_2-y_2}}{(x_2-y_2)!}\nonumber\\
+&\sum_{l=0}^\infty \frac{t^{x_2-y_1}}{(x_2-y_1)!}
\frac{t^{x_1-y_2-l}}{(x_1-y_2-l)!}
\left(-1+\frac{x_2-y_1}{t}\right)\bigg]Q,\nonumber\\
&\quad t\to\infty.
\end{align}
This is simply the propagator corresponding to a single-species
drop-push system, multiplied by $Q$.
\section{Dynamics of the particle-type}
For simplicity, let's continue with the two-particle sector. The
probability that the first particle be of type $A_\alpha$ and the
second particle be of type $A_\beta$, regardless of their
positions, is
\begin{equation}\label{41}
P^{\alpha\beta}(t)={\sum_{x_1,x_2}}'P^{\alpha\beta}(x_1,x_2;t),
\end{equation}
where the primed summation means that the summation is on the
physical region $(x_1<x_2)$. Differentiating this, one arrives at
\begin{align}\label{42}
\dot P^{\alpha\beta}(t)=&{\sum_{x_1,x_2}}'[
P^{\alpha\beta}(x_1-1,x_2;t)+P^{\alpha\beta}(x_1,x_2-1;t)
-2P^{\alpha\beta}(x_1,x_2;t)],\nonumber\\
=&\sum_x(b^{\alpha\beta}{}_{\mu\nu}-
\delta^\alpha{}_\mu\,\delta^\beta{}_\nu) P^{\mu\nu}(x-1,x;t),
\end{align}
where in the last inequality, the boundary condition \Ref{9} has
been used. It is seen that the evolution equation of the
particle-type is not closed; it involves the probability of
finding different types of the particles in adjacent sites.

However, the complete conditional probability for large times has
the simple form \Ref{40}. For that form, the summation in \Ref{41}
is readily done, and one arrives at
\begin{equation}\label{43}
{\sum_{x-1,x_2}}' U^{\alpha\beta}{}_{\mu\nu}(\vec{x};t|\vec{y};0)=
Q^{\alpha\beta}{}_{\mu\nu},\qquad t\to\infty.
\end{equation}
Here the fact has been used that the multiplier of $Q$ in \Ref{4}
is simply the propagator of the single-species drop-push model,
and its summation on the physical region results in 1. From this,
it is seen that the large-time probability of particle-types,
depends only on the initial types of the particles, and not on
their initial positions. This is of course true, when the only
eigenvalues of $b$ with modulus 1 is 1, the condition for \Ref{40}
to hold. If moreover, this eigenvalue is nondegenerate, then the
large-time probability of particle-types is even independent of
the initial particle types. In this case, $Q$ would be written
like
\begin{equation}\label{44}
Q^{\alpha\beta}{}_{\mu\nu}=q^{\alpha\beta}\,s_\mu\,s_\nu,
\end{equation}
from which
\begin{equation}\label{45}
\lim_{t\to\infty}P^{\alpha\beta}(t)=q^{\alpha\beta}.
\end{equation}
Here, $q$ is the eigenvector of $b$ with eigenvalue 1, normalized
as
\begin{equation}\label{46}
s_\alpha\,s_\beta\,q^{\alpha\beta}=1.
\end{equation}
It is seen that in this case, the large time probability of the
particle types depends only on their interaction when they are
adjacent.
\section{Concluding remarks}
It was seen that a special class of multi-species drop-push models
are solvable in the sense of the Bethe-ansatz. The condition
corresponding to this solvability, resembles very much to what
obtained in \cite{RK3} for the solvability of multi-species
asymmetric exclusion processes. This is not accidental, since the
behaviors of the drop-push model and the asymmetric exclusion
model on continuum are related to each other: using a Galilean
transformation, one can transform a drop-push model in which
particles diffuse to the right, to an exclusion model in which
particles diffuse to the left, \cite{RK,AKK2}. So the results
obtained in \cite{RK3}, regarding the solvability, with minor
modifications can be used here.

\end{document}